\documentclass[a4paper,12pt]{article}
\usepackage{graphicx,epsfig}

\begin{document}

\title{Uniformly accelerated observer in a thermal bath}
\author{Sanved Kolekar\footnote{sanved@iucaa.ernet.in}\\
IUCAA, Pune University Campus, Ganeshkhind,\\
Pune 411007, India.
}

\date{\today}
\maketitle
\begin{abstract}

We investigate the quantum field aspects in flat spacetime for an uniformly accelerated observer moving in a thermal bath. In particular, we obtain an exact closed expression of the reduced density matrix for an uniformly accelerated observer with acceleration $a = 2\pi T$ when the state of the quantum field is a thermal bath at temperature $T^\prime$. We find that the density matrix has a simple form with an effective partition function $Z$ being a product, $Z = Z_T Z_{T^\prime}$, of two thermal partition functions corresponding to temperatures $T$ and $T^\prime$ and hence is not thermal, even when $T = T^\prime$. We show that, even though the partition function has a product structure, the two thermal baths are, in fact, interacting systems; although in the high frequency limit $\omega_k \gg T$ and $\omega_k \gg T^\prime$, the interactions are found to become sub-dominant. We further demonstrate that the resulting spectrum of the Rindler particles can be interpreted in terms of spontaneous and stimulated emission due to the background thermal bath. The density matrix is also found to be symmetric in the acceleration temperature $T$ and the thermal bath temperature $T^\prime$ indicating that thermodynamic experiments alone cannot distinguish between the thermal effects due to $T$ and those due to $T^\prime$. The entanglement entropy associated with the reduced density matrix (with the background contribution of the Davies-Unruh bath removed) is shown to satisfy, in the $\omega_k \gg T^\prime$ limit, a first law of thermodynamics relation of the form $T \delta S = \delta E$  where $\delta E$ is the difference in the energies corresponding to the reduced density matrix and the background Davies-Unruh bath. The implications are discussed.

\end{abstract}

\section{Introduction}

It is well known that an uniformly accelerating observer in a flat spacetime sees the Minkowski vacuum state of a quantum field to be thermally populated with temperature $T = a/(2\pi)$ where $a$ is the magnitude of the observer's acceleration \cite{Davies, Unruh}. In fact, the reduced density matrix corresponding to Minkowski vacuum state obtained by ignoring the quantum degrees of freedom  hidden behind the casual horizon, has a form identical to that of a thermal density matrix with temperature $T = a/(2\pi)$. 

It is then interesting to ask, what does the uniformly accelerated observer see when the state of the quantum field is itself in a thermal state with some different temperature $T^\prime$? In particular, what is the form of the reduced density matrix when the \textit{state} of the quantum field is a thermal bath at temperature $T^\prime$ given the magnitude of acceleration of the accelerating observer is $a = 2\pi T$? (see also \cite{earlierworks})

One would like to know whether the reduced density matrix, in this case too, has a thermal form or atleast, in some frequency domain allows us to define an effective temperature $T_{eff}(T,T^\prime)$ for the resulting system. If so, then it would be further interesting to investigate the relationship between $T_{eff}$ and the two temperatures $T$ and $T^\prime)$; whether it is simply linearly related as $T + T^\prime$ or as  $T^2 + (T^\prime)^2$, etc which may allow one to construct a notion of a thermodynamic principle of equivalence. This remark is with keeping in view the geometrical result of \cite{deser} where it was shown that, for de-sitter spacetime, the resulting temperature detected by a constantly accelerating detector is a simple Pythagorean sum of the acceleration temperature and the de-sitter temperature. Also, for the special case $T = T^\prime$, one would like to investigate whether the two thermal baths corresponding to $T$ and $T^\prime$ equilibriate with each other or does one see physical effects due to deviations from thermal equilibrium, in particular, what is the frequency distribution of Rindler particles for such a non-equilibrium case? 

Another related issue is the following: consider a hot object, say a cup of tea, moving on an inertial trajectory in a flat spacetime. Let the object have some temperature $T^\prime$, entropy $S^\prime$ and average thermal energy $E^\prime$ measured with respect to the inertial frame.  Since the trajectory is inertial, it will eventually cross a ${\cal X} = {\cal T}$ null surface corresponding to some Rindler observer with acceleration say $a = 2\pi T$. This Rindler observer will also associate an entropy $S$ and energy $E$ with the object which are different from the quantities defined in the inertial frame due to the difference in notion of particle content for the inertial and accelerated observer. For example, in the case of the Minkowski vacuum state of the quantum field, it is known that the Rindler observer measures a non-zero entanglement entropy (assuming divergences are regulated) of the (thermal) field whereas for the inertial observer, the entropy is identically zero corresponding to the pure state. Further, from the perspective of the Rindler observer, when the hot object crosses the Rindler horizon  (or approaches a Planck length distance to the horizon), he will associate a loss of entropy with this process. It has been argued earlier (\cite{marolf}, section 4.4 of \cite{paddyinsights}) that this loss of entropy is related to the energy of the object as $S = E/T$ where $T$ is the Unruh temperature of the Rindler horizon (and not the temperature of the object itself !). However, a full quantum derivation of this result is still unknown. 

We address these issues in this paper. A related aspect pertaining to the indistinguishability of thermal and quantum fluctuations was  discussed in a separate paper \cite{san}. Here, we present a full detailed derivation of the reduced density matrix corresponding to an uniformly accelerated observer moving with acceleration $a = 2\pi T$ in a thermal bath with temperature $T^\prime$ and discuss further various physical aspects associated with the result. The paper is organized as follows. In section 2, we calculate the required reduced density matrix and show that it has a simple form given by
\begin{eqnarray}
\rho_k(n) = Z_k^{-1}(1 - Z_k^{-1})^n
\end{eqnarray}
where $n$ is the number density. The effective partition function $Z_k$ has a simple product structure given in terms of two thermal partition functions with temperatures $T^\prime$ and $T$ as $Z_k = Z_k^T Z_k^{T^\prime}$. Hence, we find that the reduced density matrix is not thermal even when $\beta = \beta^\prime$. We argue that, even though we obtain a product structure, the two thermal baths cannot be considered to be non-interacting with each other; although in the high frequency limit $\omega_k \gg T$ and $\omega_k \gg T^\prime$, we do find that the interactions become sub-dominant. We further show that the resulting spectrum of the Rindler particles can be interpreted in terms of spontaneous and stimulated emission due to the background thermal bath. The reduced density matrix is found to be symmetric under the exchange of the acceleration temperature $\beta^{-1}$ and the thermal bath temperature $(\beta^\prime)^{-1}$. This shows that, within the thermodynamic domain, Rindler observers cannot distinguish between the effects due to the thermal bath temperature from those due to the acceleration temperature. In section 3, we compute the entanglement entropy associated with the reduced density matrix (with the background contribution removed) and show that it satisfies, in the $\beta^\prime \omega \gg 1$ regime, a first law of thermodynamics relation of the form $\delta S = \beta \delta E$  where $\delta E$ is the difference in the energies corresponding to the reduced density matrix and the background Unruh bath. The conclusions are discussed in section 4.

\section{Rindler Observer in a thermal bath}

In this section, we obtain an exact closed expression of the reduced density matrix for an uniformly accelerated observer with acceleration $a = 2\pi T$ when the state of the quantum field is a thermal bath at temperature $T^\prime$.

\subsection{The setup: Reduced density matrix formalism}

Below we briefly summarize the basic idea involved. We consider the quantum field to be a scalar field obeying the dynamics of the Klein Gordon equation. Given a \textit{state} $\Phi_M$ of the scalar field  one can define a density matrix $\rho_M = |\Phi_M \rangle \langle \Phi_M | $ with which one can calculate the expectation value required of any given physical observable. Instead of a pure state, one could also have the system described by a mixed state in which case the density matrix is given as $\rho_M =\sum_i f_i |\phi_i \rangle \langle \phi_i |$. One example, relevant for our purpose here, is a mixed state defining a thermal bath. Once the $\rho_M$ of a system is known, one could in principle infer all the statistical properties of the system such as expectation values of physical observables, their mean square deviation (all higher moments as well), etc. Thus, the knowledge of $\rho_M$ is sufficient to solve the physical problem at hand. \\ \\
Further, one has the freedom to express or expand the density matrix $\rho_M $ in terms of any complete set of basis vectors. The eigenstates of the Minkowski Hamiltonian ${\cal H_{M}}$ denoted as $|_M n \rangle$ is one such natural basis. Let us denote the annihilation and creation operators corresponding to the positive  and negative Minkowski planes wave modes as $a_{k}$ and $a^\dagger_{k}$ respectively. The eigenstates corresponding to both the left and right Rindler Hamiltonian together, also form another basis eigenstates. We denote $|_L n \rangle$ and $|_R n \rangle$ to be these eigenstates respectively and $(b^\dagger_{(L)k},b_{(L)k})$ and $(b^\dagger_{(R)k},b_{(R)k})$ to be the creation and annihilation operators corresponding to the left and right Rindler positive  negative frequency modes respectively. Next, consider an observer whose time-like trajectory is restricted completely inside the right Rindler wedge ${\cal X} >|{\cal T}|$. He will have his physical observables, denoted by  ${\cal O}(b^\dagger_{(R)k},b_{(R)k})$, depend only on the right Rindler operators $(b^\dagger_{(R)k},b_{(R)k})$ and independent of the left Rindler operators $(b^\dagger_{(L)k},b_{(L)k})$. The expectation value of ${\cal O}$ for a given state of the field described by $\rho_M$ is then
\begin{eqnarray}
\langle {\cal O} \rangle &=& Tr(\rho_M {\cal O}) \nonumber \\
&=& \sum_{p,q}\langle p|_R \otimes \langle q|_L \; (\rho_M {\cal O})\; |_L q\rangle  \otimes |_R p\rangle \nonumber \\
&=& \sum_{p}\langle p|_R \; (\rho_r {\cal O}) \; |_R p\rangle
\label{expectation}
\end{eqnarray}
where we have used the independence of ${\cal O}$ on the left Rindler operators $(b^\dagger_{(L)k},b_{(L)k})$ and defined the reduced density matrix as 
 \begin{eqnarray}
\rho_r &=& \sum_{q}\langle q|_L \; \rho_M \; |_L q\rangle
\end{eqnarray}
It is evident from Eq.[\ref{expectation}] that the knowledge about the reduced density matrix $\rho_r$ is sufficient to determine the relevant expectation values in the right Rindler wedge. Following an identical procedure, one can similarly define a reduced density matrix for an observer restricted to the left Rindler wedge by tracing over the eigenstates of the right Rindler Hamiltonian. The reduced density matrix formalism applies to all those cases where certain observables depend only on a subset of total degrees of freedom of the system and in the present context, we have the casual horizon at ${\cal X} = {\cal T}$ or ${\cal X} = -{\cal T}$ acting as a natural cutoff boundary for observers restricted to the right or left Rindler wedges respectively.

\subsection{Reduced density matrix for the vacuum state}

For the Minkowski vacuum state we have $\rho_M = | 0_M\rangle \langle 0_M|$. The corresponding reduced density matrix $\rho_r$ for the right (or the left) Rindler observer obtained by tracing over the left (or the right) Rindler eigenstates turns out to be a thermal density matrix with temperature $\beta^{-1} = a/2\pi$. This is the well known Unruh effect \cite{Unruh}. The form of the thermal reduced density matrix is as follows
\begin{eqnarray}
\rho_{unruh} &=& \prod_k C_k^2 \sum_n e^{-n \beta \omega_k} |_R n_k\rangle \langle n_k|_R \nonumber \\
&=& \prod_k C_k^2 \sum_n e^{-n  \beta \omega_k} \frac{(b^\dagger_{(R)k})^n}{\sqrt{n!}} | 0_R\rangle \langle 0_R| \frac{(b_{(R)k})^n}{\sqrt{n!}}
\label{Unruhbath}
\end{eqnarray}
where $C_k^2$ is the normalization constant. The number density of particles in the above distribution is the well known Planckian
\begin{eqnarray}
\langle {\cal N}_k \rangle &=& Tr[\rho_{unruh}b^\dagger_{(R)k}b_{(R)k}] \nonumber \\
&=& \frac{1}{\left( e^{\beta \omega_k} -1 \right)}
\end{eqnarray}
In the next section, we consider the case when $\rho_M$ is a thermal density matrix.

\subsection{Reduced density matrix for a thermal bath}
 
Our aim here is to obtain the form of the reduced density matrix $\rho_r$ corresponding to the left (or the right) Rindler observer when the scalar field is described by a thermal density matrix with a temperature $ (\beta^\prime)^{-1}$. For this purpose, one would ideally like to define the density matrix for the scalar field to be given by $\rho_M = \prod_k   e^{- \beta^\prime {\cal H}_k}$ where ${\cal H}_k$ is the Minkowski Hamiltonian, ${\cal H}_k = \omega_k a^\dagger_{k}a_{k}$. Next one would like to expand $\rho_M$ completely in terms of the left and right Rindler operators and eigenstate basis. The final step would be to take a trace of the left (or the right) eigenstates and obtain the reduced density matrix.

However, this process turns out to be algebraically untractable and obtaining an exact closed form of the reduced density matrix is not possible. This is due to the well known mode mixing property of the Minkowski operators. For example the annihilation operator $a_{k}$ corresponding to a single Minkowski mode $k$ is a function of left and right Rindler operators at all Rindler frequencies, that is, $a_{k} = f\left(b_{k_1},b_{k_2},b_{k_3},...,b^\dagger_{k_1},b^\dagger_{k_2},...\right)$ and hence, a Minkowski plane wave mode of frequency $\omega_k$ mixes with Rindler plane wave modes of all frequencies. 

To avoid the issue of mode mixing, we exploit a trick due to Unruh \cite{Unruh}. Instead of the usual Minkowski plane wave modes, we work with Unruh modes $U^{(1)}_k, U^{(2)}_k$ \cite{Unruh, qfttext}. These are constructed as particular linear combinations of the usual Minkowski plane wave modes. There are two sets of operators $d^\dagger_{1k}, d_{1k}$ and $d^\dagger_{2k}, d_{2k}$ corresponding to the these Unruh modes $(U^{(1)}_k, U^{\dagger(1)}_k)$ and their complex conjugates $(U^{(2)}_k, U^{\dagger(2)}_k)$.

The Unruh modes have the following two important properties which make the present problem tractable. First, both $U^{(1)}_k$ and $U^{(2)}_k$ are linear combination of only the positive frequency Minkowski plane wave modes and hence their complex conjugates, $U^{\dagger(1)}_k$ and $U^{\dagger(2)}_k$, are a linear combination of the negative frequency Minkowski plane wave modes. Due to this, the positive and negative modes do not mix with each other and hence the annihilation operators corresponding to positive frequency Minkowski modes as well as the positive frequency Unruh modes annihilate the same state, vis-a-vis, the Minkowski vacuum state.  Second, consider an Unruh mode of frequency $k$. When expanded in terms of the left and right Rindler modes, it depends only on the $k$ and $-k$ Rindler frequencies, thus avoiding the mode mixing problem.  
      
We proceed by choosing one of the two set of Unruh operators, say $d^\dagger_{1k}, d_{1k}$, and construct the thermal density matrix as $\rho_{th} = e^{-\beta^\prime {\cal H}}$ where ${\cal H} = \sum_k \omega_k d^\dagger_{1k}d_{1k} $. Here, one should note that ${\cal H}$ is not the usual Hamiltonian ${\cal H}_{\cal M}$ of the quantum field but rather an effective Hamiltonian for a subsystem consisting of only the Unruh particles corresponding to the operators $d^\dagger_{1k}, d_{1k}$. However, since the vacuum state for ${\cal H}$ is same as that for ${\cal H}_{\cal M}$, the thermal bath constructed consists of particles created by excitations over the usual Minkowski vacuum state. In terms of the Minkowski basis eigenstates, $\rho_{th}$ is defined to be  
\begin{eqnarray}
\rho_{th} = \prod_k B_k^2 \sum_m e^{-m  \beta^\prime \omega_k} \frac{(d^\dagger_{1k})^m}{\sqrt{m!}} | 0_M\rangle \langle 0_M| \frac{(d_{1k})^m}{\sqrt{m!}}
\label{thermalbath}
\end{eqnarray}
where $B_k^2$ is the normalization constant. Further, comparing the above form of the density matrix with that of the thermal density matrix in Eq.[\ref{Unruhbath}], we can see that $ \rho_{th} $ defined is a thermal density matrix with the parameter $(\beta^\prime)^{-1}$ interpreted as the temperature of the thermal bath. 
Here, instead of the thermal bath of Unruh particles corresponding to  $d^\dagger_{1k}, d_{1k}$, we could have constructed $\rho_{th}$ in terms of ${\cal H}^\prime = d^\dagger_{2k}d_{2k}$ as well. However,  since the subsystems corresponding to ${\cal H}$ and ${\cal H}^\prime$ are different and  asymmetric in terms of the left and right Rindler modes (since $d^\dagger_{1k}$ and $d^\dagger_{2k}$ are, by definition, asymmetric in terms of the left and right Rindler creation and annihilation operators), one has to trace over different set of left or right modes in each case to avoid tracing over the thermal bath itself. For example, in the case of ${\cal H}$, one must trace over the right modes since $d^\dagger_{1k}$ depends on the left Rindler creation operator and the right Rindler annihilation operator. Similarly, in the case of ${\cal H}^\prime $, one must trace over the left Rindler modes. 
 
The choice of our $\rho_{th}$ also avoids the non-stationarity issues one encounters in the case of a thermal bath of Minkowski particles. It is known that the detector response of an Unruh-deWitt detector coupled to a $\rho^{th}_M = e^{-\beta^\prime {\cal H}_{\cal M}}$ is not a constant and is a function of the proper time along the trajectory \cite{matsas}. This due to the fact that the commutator $\left[ H_{b_L}, \rho^{th}_M \right] \neq 0$, where $H_{b_L}$ is the left Rindler Hamiltonian, indicating that unlike the Minkowski vacuum state, $\rho_M^{th}$ is not invariant under time translations along the boost Killing vector. Whereas, in the present case, we have $ \left[H_{b_L}, \rho_{th} \right] =0$ which shows that the resulting reduced density matrix is stationary in the accelerated frame.

\subsubsection{The calculation}

We next proceed to determine the reduced density matrix $\rho_r$ corresponding to $\rho_{th}$ in Eq.[\ref{thermalbath}]. To do this, we first express $\rho_{th}$ completely in terms of the left and right Rindler basis eigenstates and operators. Our final aim will be to take the trace of $\rho_{th}$ over the right Rindler eigenstates (calculation of $\rho_r$ for the right Rindler observer follows in a similar manner). We begin by expressing the Unruh operator $d^\dagger_{1k}$ as 
\begin{eqnarray}
d^\dagger_{1k} = \frac{\left[ b^\dagger_{(L)k} - {\bar q} \ b_{(R)(-k)} \right]}{{\bar p}\left(1 - {\bar q} \right)}
\label{d1intermsofb}
\end{eqnarray}
where ${\bar q} = \exp{(-\beta \omega_k/2)}$ and ${\bar p}^2 = \exp(\beta \omega_k/2)/[2\sinh{(\beta \omega_k/2)}]$. 
Using the basis representation of the left and right Rindler eigenstates, the Minkowski vacuum state can be expressed as 
\begin{eqnarray}
| 0_M\rangle = \prod_k A_k^2 \; \sum_n e^{-\frac{n}{2} \beta \omega_k} \;\; |_L n\rangle  \otimes |_R n\rangle
\label{minvacuum}
\end{eqnarray}
where $A_k^2$ is the normalization constant.
Substituting the above expansion in Eq.[\ref{thermalbath}] for the thermal density matrix $\rho_{th}$, we get
\begin{eqnarray}
\rho_{th} = \prod_k B_k^2 A_k^2 \sum_{m,p,q} e^{-m \beta^\prime \omega_k}  e^{-\frac{(p+q}{2} \beta \omega_k} \frac{(d^\dagger_{1k})^m}{m!} |_L p\rangle  \otimes |_R p\rangle \langle q|_R \otimes \langle q|_L (d_{1k})^m
\label{thermalbath2}
\end{eqnarray}
Next using the commutativity of the left and right Rindler operators 
\begin{eqnarray}
\left[(b^{(\, , \dagger)}_{(L)k},b^{(\, , \dagger)}_{(R)k}) \right] = 0
\end{eqnarray} 
we binomially expand $(d^\dagger_{1k})^m$ to obtain
\begin{eqnarray}
\left( d^\dagger_{1k}\right)^m = \frac{\left[-{\bar q}\right]^{m-l}}{\left[{\bar p}\left(1 - {\bar q} \right)\right]^m} \sum_{l=0}^m \, ^mC_l \, (b^\dagger_{(L)k})^l  \, (b_{(R)(-k)})^{m-l}
\end{eqnarray}
Substituting the above expression in Eq.[\ref{thermalbath2}], $\rho_{th}$ can be completely expressed in terms of the left and right Rindler basis eigenstates and operators as
\begin{eqnarray}
\rho_{th} &=& \prod_k C_k^2 \sum_{m,p,q,l,l^\prime} \left[\frac{e^{-m  \beta^\prime \omega_k}  e^{-\frac{(p+q}{2} \beta \omega_k}\left[-{\bar q}\right]^{2m-l-l^\prime}}{\left[{\bar p}\left(1 - {\bar q} \right)\right]^{2m}}\right]\nonumber \\\nonumber \\
&& \times \left[\frac{\, ^mC_l \, ^mC_{l^\prime} \, ^pP_{m-l} \, ^qP_{m-l^\prime}}{m!p!q!} \right] \Theta_{p-m+l}\Theta_{q-m+l^\prime} \nonumber \\\nonumber \\
&& \times (b^\dagger_{Lk})^{ p+l}(b^\dagger_{R(-k)})^{p-m+l}|_L 0\rangle  \otimes |_R 0_{(-k)}\rangle \langle 0_{(-k)}|_R \otimes \langle 0_k|_L (b_{R(-k)})^{q-m+l^\prime} (b_{Lk})^{q+l^\prime} \nonumber
\end{eqnarray}
where $C_k^2 = B_k^2 A_k^2$ and $\Theta_x$ is the Heaviside step function. Simplifying the above expression, we get
\begin{eqnarray}
\rho_{th} &=& \prod_k C_k^2 \sum_{m,p,q,l,l^\prime} \left[\frac{e^{-m \beta^\prime \omega_k}  e^{-\frac{(p+q}{2} \beta \omega_k}\left[-{\bar q}\right]^{2m-l-l^\prime}}{\left[{\bar p}\left(1 - {\bar q} \right)\right]^{2m}}\right]\nonumber \\\nonumber \\
&& \times \left[\frac{\, ^mC_l \, ^mC_{l^\prime} \, ^pP_{m-l} \, ^qP_{m-l^\prime}}{m!p!q!} \right] \Theta_{p-m+l}\Theta_{q-m+l^\prime} \nonumber \\\nonumber \\
&& \times \sqrt{(p+l)!(q+l^\prime)!(p-m+l)!(q-m+l^\prime)!}\nonumber \\\nonumber \\
&& \times \; |_L (p+l)\rangle  \otimes |_R (p-m+l)\rangle \langle (q-m+l^\prime) |_R \otimes \langle (q+l^\prime)|_L
\label{thermalbath3}
\end{eqnarray}
It is convenient to take the trace of $\rho_{th}$ in the above form. Tracing over the right Rindler states, we get
\begin{eqnarray}
\rho_{r(left)} &=& \sum_{X}\langle X|_R \; \rho_{th} \; |_R X\rangle \nonumber \\
&=& \prod_k C_k^2 \sum_{m,p,l,l^\prime} \left[\frac{e^{-m  \beta^\prime \omega_k}  e^{-(p+l^\prime) \beta \omega_k}}{\left[{\bar p}\left(1 - {\bar q} \right)\right]^{2m}}\right] \left[\frac{(-l)^{l+l^\prime}\, ^mC_l \, ^mC_{l^\prime} \, ^pP_{l}}{m!p!} \right] \Theta_{p-l} \nonumber \\\nonumber \\
&& \times (p+m-l)! \; |_R (p-l+m)\rangle \langle (p-l+m) |_R 
\label{rightreducedM1}
\end{eqnarray}
where we have summed over $X,q$ and used the symmetry of $l \rightarrow (m-l)$ and $l^\prime \rightarrow (m-l^\prime)$. Further making the shift $p \rightarrow (p+l)$, it is possible to separate the terms involving $l$ and $l^\prime$. We get
\begin{eqnarray}
\rho_{r(left)} &=& \prod_k C_k^2 \sum_{m,p,l,l^\prime} \left[\frac{e^{-m \beta^\prime \omega_k}  e^{-p \beta \omega_k}}{p!m!}\right] \left[(-l)^{l+l^\prime}\, ^mC_l \, ^mC_{l^\prime} e^{-(l+l^\prime) \beta \omega_k} \right]  \nonumber \\
&& \times (p+m)! \; |_R (p+m)\rangle \langle (p+m) |_R 
\label{rightreducedMinbetween}
\end{eqnarray}
where we have used the definitions of $ \bar p$ and $\bar q$. We can now sum over $l$ and $l^\prime$ to get
\begin{eqnarray}
\rho_{r(left)} &=& \prod_k C_k^2 \sum_{m,p}\; ^{m+p}C_m \left[e^{- ( \beta^\prime+\beta) \omega_k}  (e^{ \beta \omega_k}-1)\right]^m  e^{ -p\beta \omega_k} \; |_R (p+m)\rangle \langle (p+m) |_R \nonumber \\
&=& \prod_k C_k^2 \sum_{p} \left[1+e^{- ( \beta^\prime-\beta) \omega_k}  (1 - e^{- \beta \omega_k})\right]^{p} e^{ -p\beta \omega_k} \; |_R p\rangle \langle p |_R
\label{leftreducedM}
\end{eqnarray}
where the normalization constant $C_k^2$ can be obtained as $C_k^2 = (1 - e^{- \beta \omega_k})(1 - e^{- \beta^\prime \omega_k})$.
One should note that the above form of $\rho_r$ is exact and does not involve any approximation.  The expectation value of the number operator give us the frequency distribution of particles seen by the left Rindler observers. We find
\begin{eqnarray}
\langle {\cal N}^ k_{left} \rangle &=& Tr[\rho_{r(right)}b^\dagger_{(R)k}b_{(R)k}] \nonumber \\
&=& \frac{1}{\left( e^{\beta \omega_k} -1 \right)} +\frac{1}{\left( e^{\beta^\prime \omega_k} -1 \right)} + \frac{1}{\left( e^{\beta^\prime \omega_k} -1 \right)\left( e^{\beta \omega_k} -1 \right)}
\label{leftexpectation}
\end{eqnarray}

\subsection{The result}

Here, it is advantageous to introduce the following two (partition) functions,
\begin{eqnarray}
(Z_k^\beta)^{-1} &=& (1 - e^{- \beta \omega_k}) \\
(Z_k^{\beta^\prime})^{-1} &=& (1 - e^{- \beta^\prime \omega_k})
\end{eqnarray}
We can recast the reduced density matrix in Eq.[\ref{leftreducedM}] in a much simpler form in terms of these functions. We get 
\begin{eqnarray}
\rho_{r(left)} &=& \prod_k (Z_k^\beta Z_k^{ \beta^\prime})^{-1} \sum_{p} \left[1 - (Z_k^\beta Z_k^{\beta^\prime})^{-1} \right]^{p} \; |_L p\rangle \langle p |_L \nonumber \\
&=& \prod_k ({\bar Z^{left}_k(\beta, \beta^\prime)})^{-1} \sum_{p} \left[1 - ({\bar Z^{left}_k(\beta, \beta^\prime)})^{-1} \right]^{p} \; |_L p\rangle \langle p |_L 
\label{leftreducedM2}
\end{eqnarray}
where we have defined $ {\bar Z_k(\beta, \beta^\prime)} $ to be $ {\bar Z_k(\beta, \beta^\prime)} = Z_k^\beta Z_k^{\beta^\prime}$. One can aslo rewrite the thermal density marix in Eq.[\ref{Unruhbath}] in term of the function $(Z_k^\beta)^{-1}$ to get
\begin{eqnarray}
\rho_{(thermal)} &=& \prod_k (Z_k^\beta)^{-1} \sum_{p} \left[1 - (Z_k^\beta)^{-1} \right]^{p} \; |_L p\rangle \langle p |_L 
\label{rhothermal}
\end{eqnarray}
From the above form of the thermal density matrix, we can read off $Z_k^\beta$ to be the partition function for a single mode of frequency $\omega_k$ corresponding to a thermal bath at temperature $\beta^{-1}$. Similarly,  $Z_k^{\beta^\prime}$ is the partition function corresponding to a thermal bath at temperature $(\beta^\prime)^{-1}$. Comparing the expression for the reduced density matrix in Eq.[\ref{leftreducedM2}] with that of the thermal form, we can see they have the same structural form $\rho = Z^{-1} \sum_{n} (1 - Z^{-1})^n \; | n\rangle \langle n |$. Further, we can read off $ {\bar Z_k(\beta,\beta^\prime)} = Z_k^\beta Z_k^{\beta^\prime}$ to be the effective partition function of $\rho_{r(left)}$. For these type of structural forms, the expectation value of the number density $N_k$ turns out to be just $N_k = Z_k-1 $. This can easily be verified for $\rho_{r(left)}$ in Eq.[\ref{leftreducedM2}]. 

There are several interesting features of the reduced density matrix  which we will list below.

\begin{enumerate}
\item The effective partition function ${\bar Z_k(\beta, \beta^\prime)}$ has a simple product structure given in terms of two thermal partition functions with temperatures $(\beta^\prime)^{-1}$ and $ \beta^{-1}$ as ${\bar Z_k(\beta, \beta^\prime)} = Z_k^\beta Z_k^{\beta^\prime}$. This by itself is a remarkable result.

\item Due to the product nature,  the reduced density matrix has an symmetry, $\rho(\beta, \beta^\prime) = \rho( \beta^\prime, \beta)$. Thus, purely within the thermodynamic domain, the Rindler observer cannot distinguish between the thermal effects due to the thermal bath temperature and those due to the acceleration temperature. This symmetry strengthens the indistinguishability of quantum and thermal fluctuations previously known in the case of vacuum states. This aspect was discussed in detail in a separate paper \cite{san}.

\item When there is no thermal bath present, that is, when $(\beta^\prime)^{-1} \rightarrow 0$, we have $Z_k^{\beta^\prime} \rightarrow 1$. We then retrieve the Davies-Unruh bath, $\rho_r \rightarrow \rho_{unruh}$ as expected. Similarly, when the acceleration $a$ goes to zero, that is, when $ \beta^{-1} \rightarrow 0$, we have $Z_k^{ \beta} \rightarrow 1$. Then we retrieve the initial thermal bath  $\rho_r \rightarrow \rho_{thermal}$, again, as expected.

\item The resulting form of $\rho_r$ is not thermal. Also, it does not seem possible to simply define an effective temperature for the system. Even, in the special case $\beta = \beta^\prime$, we have $ {\bar Z_k(\beta, \beta^\prime)}_{\beta = \beta^\prime} = (Z_k^\beta )^2$ and hence no thermality. 

\item In the regime, $\beta \omega \ll 1$ and $ \beta^\prime \omega \ll 1$, we have upto quadratic order
\begin{eqnarray}
 {\bar Z_k(\beta,  \beta^\prime)} = 1 - \beta \beta^\prime \omega_k^2
\end{eqnarray}
Thus scaling $\beta \rightarrow \alpha \beta$ and  $ \beta^\prime \rightarrow \beta^\prime/\alpha$ leaves the effective partition function unchanged upto quadratic order.

\end{enumerate}

For the sake of completeness, we give below the expression for the reduced density matrix for the right Rindler observer as well. It is obtained by tracing over the left Rindler states in a similar manner as in the left Rindler case.  
\begin{eqnarray}
\rho_{r(right)} &=& \prod_k C_k^2 \sum_{m,p}\; ^{mp+}C_m \left[e^{- (\beta^\prime+\beta) \omega_k}  (e^{ \beta \omega_k}-1)\right]^m  e^{ -p\beta \omega_k} \; |_R p\rangle \langle p |_R \nonumber \\
&=& \prod_k C_k^2 \sum_{p} \left[1-e^{- (\beta^\prime+\beta) \omega_k}  (e^{ \beta \omega_k}-1)\right]^{-p-1} e^{ -p\beta \omega_k} \; |_R p\rangle \langle p |_R
\label{rightreducedM}
\end{eqnarray}
The normalization constant $C_k^2$ is  found by the normalization condition $Tr(\rho_r) =1$ to be $C_k^2 = (1 - e^{- \beta \omega_k})(1 - e^{- \beta^\prime \omega_k})/[1 - (1 - e^{- \beta \omega_k})e^{-\beta^\prime \omega_k}]$.
In terms of the partition functions $Z_k^\beta $ and $Z_k^{ \beta^\prime} $, the $\rho_{r(right)}$ simply becomes
\begin{eqnarray}
\rho_{r(right)} &=& \prod_k (Z_k^\beta Z_k^{\beta^\prime} - Z_k^{\beta^\prime} + 1)^{-1} \sum_{p} \left[1 - (Z_k^\beta Z_k^{\beta^\prime} - Z_k^{\beta^\prime} + 1)^{-1} \right]^{p} \; |_L p\rangle \langle p |_L \nonumber \\
&=& \prod_k ({\bar Z^{right}_k(\beta, \beta^\prime)})^{-1} \sum_{p} \left[1 - ({\bar Z^{right}_k(\beta, \beta^\prime)})^{-1} \right]^{p} \; |_L p\rangle \langle p |_L 
\label{rightreducedM2}
\end{eqnarray}
The expectation value of the number operator is
\begin{eqnarray}
\langle {\cal N}_{right} \rangle &=& Tr[\rho_{r(right)}b^\dagger_{(R)k}b_{(R)k}] \nonumber \\
&=& \frac{1}{\left( e^{\beta \omega_k} -1 \right)}  + \frac{1}{\left( e^{\beta^\prime \omega_k} -1 \right)\left( e^{\beta \omega_k} -1 \right)}
\label{rightexpectation}
\end{eqnarray}
From the above expression one can see that the contribution from the thermal bath $1/\left( e^{\beta^\prime \omega_k} -1 \right)$ is absent as compared to that in Eq.[\ref{leftexpectation}] for the left Rindler observer. This is due to the asymmetric construction of the Unruh modes in terms of the left and right Rindler modes and as mentioned earlier, since in the present context, we have chosen the modes corresponding to $d^\dagger_{1k}, d_{1k}$, it is appropriate to trace over the right Rindler modes to avoid tracing over the thermal bath itself.

\section{Entanglement entropy of $\rho_r$}

We calculate the entanglement entropy associated with the reduced density matrix $\rho_r$. The entropy is given by the familiar expression
\begin{eqnarray}
\bar S = -Tr(\rho_r \ln{\rho_r})
\end{eqnarray}
Substituting Eq.[\ref{leftreducedM2}] in the above expression, we get
\begin{eqnarray}
\bar S &=& -\sum_k \ln{\bar Z_k^{-1}} - \prod_k {\bar Z_k^{-1}} \sum_p (1- {\bar Z_k^{-1}})^p p \sum_m \ln (1 - {\bar Z_m^{-1}}) \nonumber \\
 &=& -\sum_k \ln{\bar Z_k^{-1}} - \sum_k \langle {\cal N}^ k_{left} \rangle \ln (1 - {\bar Z_k^{-1}}) 
\end{eqnarray}
where we have used the definition $\langle {\cal N}^ k_{left} \rangle = \sum_p p {\bar Z_k^{-1}}  (1- {\bar Z_k^{-1}})^p$ and $Tr(\rho_k) = 1$ to arrive at the last expression. We know that the expression of entanglement entropy for vacuum state, excited states contain divergences and hence we expect the same to happen in the case of $\bar S$ as well. However, since we will be mainly interested in the difference between the entropy of two different states, we expect these divergences to cancel and lead to meaningful physical expressions. We first calculate $\bar S - S_{\beta}$ where $S_{\beta}$ is the entanglement entropy corresponding to the Davies-Unruh thermal density matrix in Eq.[\ref{Unruhbath}]. We get
\begin{eqnarray}
\Delta S_1 & = & \bar S - S_{\beta} \nonumber \\
& = &  -\sum_k \ln{\frac{\bar Z_k^{-1}}{Z_{\beta}^{-1}}} - \sum_k \langle {\cal N}^ k_{left} \rangle \ln (1 - {\bar Z_k^{-1}}) - \sum_k \langle {\cal N}^ k_{\beta} \rangle \ln (1 - {Z^\beta_k}^{-1}) \nonumber \\
& = & \beta \delta E_1 - \sum_k \ln{\left( 1 - e^{- \beta^\prime \omega_k} \right) } - \sum_k \langle {\cal N}^ k_{left} \rangle \ln{\left[ 1 + (e^{\beta \omega_k} - 1) e^{- \beta^\prime \omega_k} \right]} \nonumber \\
\label{S1}
\end{eqnarray}
where $\delta E_1 = \sum_k \omega_k \left(\langle {\cal N}^ k_{left} \rangle -  \langle {\cal N}^ k_{\beta} \rangle \right)  =Tr({\cal H}( \rho_r - \rho_{\beta})) $ is the difference between the energies associated with $\rho_r$ and $\rho_{\beta}$ respectively. The above expression is exact. Further, consider the case where $T^\prime$ is small, which is the case when the backreaction of the thermal bath on the background flat metric is negligible. In this regime, we have $\beta^\prime \rightarrow \infty$ and $e^{- \beta^\prime \omega_k} \sim x \ll 1$. We next expand Eq.[\ref{S1}] upto linear order in the parameter $x$. It is easy to check that the last two terms in Eq.[\ref{S1}] do not contribute in the linear order whereas only the first term survives. Hence, we finally get 
\begin{eqnarray}
\Delta S_1  \approx  \delta S_1  \approx   \beta \delta E_1 
\label{S1approx}
\end{eqnarray}
which is the result we wanted to prove. We find that although $\rho_r$ is not thermal, in the regime  $e^{-\beta^\prime \omega_k} \ll 1$, the differences in entropy and energy of  the two \textit{states} obey the first law of thermodynamics with the temperature involved being the Unruh temperature of the Rindler horizon.

Similarly, exploiting the symmetry of $\beta $ and $\beta^\prime$, one can show
\begin{eqnarray}
\delta S_2 & = & \bar S - S_{ \beta^\prime} \nonumber \\
& \approx &  \beta (\bar E - E_{\beta^\prime}) 
\label{S2}
\end{eqnarray} 
when  $e^{-  \beta \omega_k} \ll 1$. Physically this would correspond to the limit $T \rightarrow 0$ when the acceleration of the Rindler observer is small.
 
\section{Discussion and conclusions}

We have obtained an exact closed expression of the reduced density matrix for an uniformly accelerated observer with acceleration $a = 2\pi \beta^{-1}$ in a thermal bath of temperature $(\beta^\prime)^{-1}$. To summarize, we found that the reduced density matrix has a simple form given as
\begin{eqnarray}
\rho_{r} &=& \prod_k ({\bar Z_k(\beta, \beta^\prime)})^{-1} \sum_{p} \left[1 - ({\bar Z_k(\beta, \beta^\prime)})^{-1} \right]^{p} \; |_L p\rangle \langle p |_L 
\label{rhofinal}
\end{eqnarray}
The effective partition function ${\bar Z_k(\beta, \beta^\prime)}$ has a simple product structure given in terms of two thermal partition functions as ${\bar Z_k(\beta, \beta^\prime)} = Z_k^\beta Z_k^{\beta^\prime}$. It is evident that $\rho_r$ is not a thermal density matrix, even for the special case of $\beta = \beta^\prime$ and moreover, due to the product structure, there is no simple way to combine the two temperatures or even define an effective temperature for the system. However, due to the product structure,  the reduced density matrix is found to have an additional symmetry $\rho(\beta,  \beta^\prime) = \rho( \beta^\prime, \beta)$.  

Below, we discuss the various features related to $\rho_r$. First, we begin by looking at the functional form of the resulting density matrix. It can be simply expressed as 
\begin{eqnarray}
\rho_{ \{ n \}}  = Z^{-1} (1- Z^{-1})^n
\label{Jaynesrho}
\end{eqnarray} 
The above form of the distribution obeys the simplest version of the principle of maximal entropy wherein one extremizes the entropy $S$ along with all the information which is available \cite{jayne, bek}. This can be demonstrated as follows: In most of the simplest cases, one has the following three inputs. First, the entropy of the system given by $S = -Tr(\rho \ln{\rho})$. Second the normalization condition $Tr(\rho) = 1$ and third, $Tr(\rho {\hat n}) = \langle n \rangle$ where the expectation value $\langle n \rangle$ is given or prescribed (Instead of $\langle n \rangle$, one could also be given $\langle E \rangle$ = $\omega \langle n \rangle$, the average energy of the system). Then the maximal entropy principle demands the following to hold.
\begin{eqnarray}
\delta \left( -\sum_{ \{ n \}} \rho_{ \{ n \}} \ln{\rho_{ \{ n \}}} + C_1 \sum_{ \{ n \}}  n \rho_{ \{ n \}}  + C_2 \sum_{ \{ n \}}  \rho_{ \{ n \}}  \right) = 0
\end{eqnarray}
where $C_1$ and $C_2$ are the Lagrange multipliers corresponding to the two constraint equations. It can be easily verified that $\rho$ of the form in Eq.[\ref{Jaynesrho}] satisfies the above extremum where one has fixed one of Lagrange multipliers using the normalization condition $Tr(\rho) = 1$ and the second Lagrange multiplier has been expressed in terms of $\langle n \rangle$ which is assumed to be given. The expectation value $\langle n \rangle$ can further be expressed in terms of the effective partition function $Z$ through the relation $\langle n \rangle = Z-1 $ which holds for the form of the $\rho$ in Eq.[\ref{Jaynesrho}].

Second, the effective partition function we have obtained is a product of two thermal partition functions at different inverse temperatures, $\beta$ and $ \beta^\prime$. Usually, in the study of  statistical mechanics of a system, one encounters a similar product in the case of non-interacting sub-systems in equilibrium with a thermal bath at a common temperature $\beta$. For example, consider a system of $N$ non-interacting particles in thermal equilibrium. The total partition function for the system is simply given as
\begin{eqnarray}
Z_T &=& Tr(e^{-\beta {\cal H}_1 - \beta {\cal H}_2 - \cdots}) \nonumber \\
&=& Z_1 Z_2 \cdots Z_N
\label{statpart}
\end{eqnarray}
In the present context, one may be tempted to conclude that the thermal bath with partition function $Z_k^{\beta^\prime}$  and the Davies-Unruh bath with partition function $Z_k^\beta$ are not interacting with one another and each bath maintains its own thermal equilibrium at its respective temperature. However, this would be incorrect since from the form of the number density of particles, $\langle {\cal N}^ k_{left} \rangle$ in Eq.[\ref{leftexpectation}], one can see that it contains a term involving a product of planckians with inverse temperature $\beta$ and $\beta^\prime$. This extra term signifies that even though the effective partition function is a product,  there are  quantum correlation present between the two thermal baths. This is to be expected since the tracing operation non-trivially entangles the left and right modes thereby introducing correlations between the two baths. Further, it is obvious that the reduced density matrix cannot be expressed in the form similar to Eq.[\ref{statpart}], that is $\rho \neq e^{-\beta {\cal H}_1 - \beta {\cal H}_2}$ (when the two temperatures are equal) which is required for the sub-systems to be non-interacting systems. Instead, from Eq.[\ref{leftreducedM2}], we see that $\rho_r$ is of the form $\rho_r \propto (1-Z_1 Z_2)^{\hat n}$ which essentially mixes the two baths.

Third, let us next define $n_k^{\beta} = 1/(e^{\beta \omega_k} -1)$ and $n_k^{\beta^\prime} = 1/(e^{\beta^\prime \omega_k} -1)$. In terms of these, the expectation value $\langle {\cal N}^ k_{left} \rangle$ in Eq.[\ref{leftexpectation}] can be written as
\begin{eqnarray}
\langle {\cal N}^ k_{left} \rangle &=& n_k^{\beta} +n_k^{\beta^\prime} + n_k^{\beta^\prime} n_k^{\beta} \nonumber \\
& = & n_k^{\beta^\prime} + n_k^{ \beta} \left( n_k^{ \beta^\prime} + 1 \right)
\end{eqnarray}
The above spectrum of particles can be explained in the usual way (see for example, ref \cite{sponstim}) in terms of spontaneous and stimulated emission of particles. Consider the $n_k^{ \beta} \left( n_k^{\beta^\prime} + 1 \right)$ term first. The $+1$ inside the bracket indicates the spontaneous emission of  $n_k^{\beta} $ in the absence of $n_k^{ \beta^\prime} $, which is just the Unruh effect, while  the product $n_k^{ \beta} n_k^{ \beta^\prime}$ signifies the stimulated emission of additional $n_k^{ \beta^\prime} $ due to the ambient Davies-Unruh bath $n_k^{\beta} $. The remaining first term $n_k^{ \beta^\prime} $ is just the distribution of particles in the original thermal bath. One can also note that $\langle {\cal N}^ k_{left} \rangle$ is symmetric in $n_k^{ \beta^\prime} $ and $n_k^\beta $. Thus, it is possible to explain the stimulated and stimulated emission processes with the roles reversed between $n_k^{ \beta^\prime} $ and $n_k^\beta $ too. Such a symmetry was pointed out earlier in  \cite{bek, sponstim} in the case of black hole immersed in a background thermal bath. However, in the present case, the reduced density matrix also has an additional symmetry in $\beta \Leftrightarrow  \beta^\prime$ which in turn implies the symmetry in $n_k^{ \beta^\prime} \Leftrightarrow n_k^\beta $.

Further, if one considers the high frequency regime where $e^{- \beta^\prime \omega_k} \sim x \ll 1$, and $ e^{- \beta \omega_k} \sim y \ll 1$ then one can neglect $n_k^{ \beta^\prime} \approx e^{- \beta^\prime \omega_k}$ compared to $1$ in the second term in the above equation. In that case, we get upto linear order in $x$ and $y$
\begin{eqnarray}
\langle {\cal N}^ k_{left} \rangle & \approx & n_k^{\beta} +n_k^{\beta^\prime} 
\end{eqnarray}
Thus, the interaction term is sub-dominant in the high frequency regime where the quantum effects become more and more local suggesting that the cross-correlations between the two thermal baths exists mostly on large scales compared to $\beta$ and $ \beta^\prime$. This conclusion is coherent with an earlier result \cite{sanbrownian} where we showed that there are anisotropies in the quantum correlations in fluctuations of the Davies-Unruh bath at scales larger compared to temperature $\beta^{-1}$ of the bath, however in the high frequency limit, these differences vanish and one recovers the classically expected isotropy in the correlations.

Finally, we conclude by mentioning that even though $\rho_r$ is not thermal, the difference in the entanglement entropies $\bar S - S_{\beta}$ corresponding to the thermal bath and the Davies-Unruh bath respectively, follow the first law of thermodynamics and is related to the difference of the respective energies $\delta E_1 =Tr({\cal H}( \rho_r - \rho_{\beta})) $ as $\delta S_1  \approx   \beta \delta E_1 $ in the $ \beta^\prime \omega_k \gg 1$ regime with the temperature $T$ being the Unruh temperature of the Rindler horizon.

\section*{Acknowledgments}
We thank T. Padmanabhan and Jorma Louko for useful discussions and comments.



\begin{thebibliography}{100}

\bibitem{Davies}
P. C. W. Davies,  J. Phys. A \textbf{8}, 609 - 616 (1975).

\bibitem{Unruh}
W. G. Unruh,  Phys. Rev. D \textbf{14}, 870 (1976).

\bibitem{earlierworks}
The response rate of the Unruh-DeWitt detector coupled to a thermal bath was investigated in; T. Padmanabhan and T.P. Singh, Phys. Rev. D \textbf{38}, 2457-2463 (1988); S. S. Costa and G. E. A. Matsas, Phys. Rev. D \textbf{52},  3466-3471 (1995). However, one should note that these are not first principle derivations and contain prescriptions which involve taking suitable thermal weightages of different excited states.



\bibitem{deser} 
S. Deser and O. Levin, Class. Quant. Grav. \textbf{14}, L163-L168 (1997).

\bibitem{marolf}
D. Marolf, D. Minic and S. Ross, Phys.Rev. D \textbf{69}, 064006
(2004).

\bibitem{paddyinsights}
T.Padmanabhan, Rep. Prog. Phys. \textbf{73}, 04690 (2010) [arXiv:0911.5004].

\bibitem{san} 
Sanved Kolekar and T. Padmanabhan, \textit{Indistinguishability of thermal and quantum fluctuations} [arXiv:1308.6289].

\bibitem{qfttext}
Bryce DeWitt, The Global Approach to Quantum Field Theory (Clarendon Press, Oxford, 2003); N. D. Birrell and P. C. W. Davies, Quantum Fields in Curved
Space (Cambridge University Press, Cambridge, England, 1982).


\bibitem{matsas} 
S. S. Costa and G. E. A. Matsas, Phys. Rev. D \textbf{52},  3466-3471 (1995).

\bibitem{jayne}
E. T. Jaynes, Phys. Rev. \textbf{106}, 620 (1957).

\bibitem{bek}
J. D. Bekenstein , Phys. Rev. D \textbf{12}, 3077–3085 (1975).

\bibitem{sponstim}
R. Wald, Phys. Rev. D \textbf{13}, 3176–3182 (1976); J. D. Bekenstein and A Meisels , Phys. Rev. D \textbf{15}, 2775–2781 (1977).

\bibitem{sanbrownian}
Sanved Kolekar and T. Padmanabhan, Phys. Rev. D \textbf{86}, 104057 (2012) [arXiv:1205.0258].



  
\end{thebibliography}
\end{document}